\RequirePackage[hyphens]{url}
\documentclass[prd,twocolumn,a4paper,floatfix,nofootinbib,preprintnumbers,superscriptaddress]{revtex4-1}
\usepackage[utf8]{inputenc}
\usepackage{hyperref}
\usepackage[nolist,nohyperlinks]{acronym}
\usepackage{amsthm}
\usepackage{fontawesome5}
\usepackage{graphicx}
\usepackage{amsfonts}
\usepackage{booktabs}
\usepackage{siunitx}
\usepackage{mathtools}
\usepackage{multirow}
\usepackage{array}
\usepackage[normalem]{ulem}
\usepackage{orcidlink}
\usepackage{natbib}
\usepackage{bm}
\usepackage{cleveref}
\usepackage{xcolor}
\usepackage{caption}
\usepackage{float}
\usepackage[nolist,nohyperlinks]{acronym}
\usepackage{tabularx}
\usepackage{physics}
\usepackage{subcaption}
\usepackage{placeins}
\usepackage[justification=RaggedRight]{caption}
\usepackage{transparent}
\usepackage[multiple]{footmisc}
\usepackage[dont-mess-around]{fnpct}
\usepackage{import}
\newcommand*\diff{\mathop{}\!\mathrm{d}}

\newcommand{\thetaeos}{\boldsymbol{\theta}_{\rm{EOS}}}
\newcommand{\boldtheta}{\boldsymbol{\theta}}
\newcommand{\MTOV}{M_{\rm{TOV}}}
\newcommand{\nbreak}{n_{\rm{break}}}
\newcommand{\nsat}{n_{\rm{sat}}}
\newcommand{\chiEFT}{\chi_{\rm{EFT}}}

\begin{document}

\title{Leveraging differentiable programming in the inverse problem of neutron stars}

\author{Thibeau Wouters\thanks{t.r.i.wouters@uu.nl} \orcidlink{0009-0006-2797-3808}}
\email{t.r.i.wouters@uu.nl}
\affiliation{Institute for Gravitational and Subatomic Physics (GRASP), Utrecht University, Princetonplein 1, 3584 CC Utrecht, The Netherlands}
\affiliation{Nikhef, Science Park 105, 1098 XG Amsterdam, The Netherlands}
\author{Peter T. H. Pang \orcidlink{0000-0001-7041-3239}}
\affiliation{Nikhef, Science Park 105, 1098 XG Amsterdam, The Netherlands}
\affiliation{Institute for Gravitational and Subatomic Physics (GRASP), Utrecht University, Princetonplein 1, 3584 CC Utrecht, The Netherlands}
\author{Hauke \surname{Koehn}\,\orcidlink{0009-0001-5350-7468}}
\affiliation{Institut f\"ur Physik und Astronomie, Universit\"at Potsdam, Haus 28, Karl-Liebknecht-Str. 24/25, 14476, Potsdam, Germany}
\author{Henrik \surname{Rose}\,\orcidlink{0009-0009-2025-8256}}
\affiliation{Institut f\"ur Physik und Astronomie, Universit\"at Potsdam, Haus 28, Karl-Liebknecht-Str. 24/25, 14476, Potsdam, Germany}
\author{Rahul \surname{Somasundaram}\,\orcidlink{0000-0003-0427-3893}}
\affiliation{Department of Physics, Syracuse University, Syracuse, NY 13244, USA}
\affiliation{Theoretical Division, Los Alamos National Laboratory, Los Alamos, NM 87545, USA}
\author{Ingo \surname{Tews}\,\orcidlink{0000-0003-2656-6355}}
\affiliation{Theoretical Division, Los Alamos National Laboratory, Los Alamos, NM 87545, USA}
\author{Tim Dietrich \orcidlink{0000-0003-2374-307X}}
\affiliation{Institut f\"ur Physik und Astronomie, Universit\"at Potsdam, Haus 28, Karl-Liebknecht-Str. 24/25, 14476, Potsdam, Germany}
\affiliation{Max Planck Institute for Gravitational Physics (Albert Einstein Institute), Am M\"{u}hlenberg 1, Potsdam 14476, Germany}
\author{Chris Van Den Broeck \orcidlink{0000-0001-6800-4006}}
\affiliation{Institute for Gravitational and Subatomic Physics (GRASP), Utrecht University, Princetonplein 1, 3584 CC Utrecht, The Netherlands}
\affiliation{Nikhef, Science Park 105, 1098 XG Amsterdam, The Netherlands}
\date{\today}

\preprint{LA-UR-25-23486}

\begin{abstract}
\Acp{NS} probe the high-density regime of the nuclear \ac{EOS}.
However, inferring the EOS from observations of \acp{NS} is a computationally challenging task.
In this work, we efficiently solve this inverse problem by leveraging differential programming in two ways.
First, we enable full Bayesian inference in under one hour of wall time on a GPU by using gradient-based samplers, without requiring pre-trained machine learning emulators.
Moreover, we demonstrate efficient scaling to high-dimensional parameter spaces. 
Second, we introduce a novel gradient-based optimization scheme that recovers the \ac{EOS} of a given \ac{NS} mass-radius curve.
We demonstrate how our framework can reveal consistencies or tensions between nuclear physics and astrophysics.
First, we show how the breakdown density of a metamodel description of the \ac{EOS} can be determined from \ac{NS} observations.
Second, we demonstrate how degeneracies in EOS modeling using nuclear empirical parameters can influence the inverse problem during gradient-based optimization.
Looking ahead, our approach opens up new theoretical studies of the relation between \ac{NS} properties and the \ac{EOS}, while effectively tackling the data analysis challenges brought by future detectors. 
\end{abstract}

\maketitle

\acresetall

\section{Introduction}\label{sec:intro}

The nuclear \ac{EOS} remains uncertain and an open problem in physics~\cite{Lattimer:2012nd, Ozel:2016oaf, Burgio:2021vgk, Chatziioannou:2024tjq}. 
Nuclear experiments only constrain its lower density behavior, i.e., around the nuclear saturation density, ${\rho_{\rm{sat}} \equiv 2.7 \times 10^{14} \ \rm{g} \ \rm{cm}^{-3}}$ (or equivalently, a baryon number density of ${n_{\rm{sat}} \equiv \SI{0.16}{\per\femto\meter\cubed}}$).
For instance, experiments measuring the neutron-skin thickness of atomic nuclei constrain the symmetry energy at nuclear saturation~\cite{Birkhan:2016qkr, Thiel:2019tkm, PREX:2021umo, Reed:2021nqk, CREX:2022kgg, Giacalone:2023cet}, while heavy-ion collisions uniquely probe isospin-symmetric matter above the nuclear saturation density~\cite{Danielewicz:2002pu,Li:2008gp,Huth:2021bsp,Sorensen:2023zkk}. 
In addition, nuclear theory calculations provide constraints on isospin-asymmetric matter up to two times the nuclear saturation density~\cite{Tews:2018kmu,Drischler:2020hwi,Armstrong:2025tza}.

Matter at the most extreme densities, up to $6-8 n_{\rm{sat}}$, is only realized inside \acp{NS}, the remnants of core-collapse supernovae~\cite{Lattimer:2004sa}. 
Their macroscopic properties, such as mass, radius, quadrupole moment, and tidal deformability, are determined through the \ac{TOV} equations, which require the \ac{EOS} for closure.
Consequently, measurements of \ac{NS} properties can then be used to infer the \ac{EOS}~\cite{Lindblom1992-zb}, which is referred to as the inverse problem of \acp{NS}. 
Several such measurements are now available. 
For instance, precise radio timing of binary pulsar systems determines the masses of \acp{NS}~\cite{Hobbs:2006cd, Demorest:2010bx}. 
Furthermore, pulse-profile modeling of X-ray emissions from hotspots of millisecond pulsars provides joint estimates on the mass and radius of \acp{NS}~\cite{Watts:2019lbs}, and has been achieved with the \ac{NICER}~\cite{gendreau2016neutron, Bogdanov:2019ixe, riley2023xpsi, Riley:2019yda, Miller:2019cac, Riley:2021pdl, Miller:2021qha}.
Additionally, the tidal deformability of \acp{NS} leaves a measurable imprint on \acp{GW} emitted during the inspiral and merger of \ac{BNS} systems.
This way, GW170817~\cite{LIGOScientific:2017vwq, LIGOScientific:2018hze}, the first \ac{BNS} merger observed by Advanced LIGO~\cite{LIGOScientific:2014pky} and Advanced Virgo~\cite{VIRGO:2014yos}, provided the first estimates on \ac{NS} radii derived from \acp{GW}~\cite{LIGOScientific:2018cki}. 

Constraining the \ac{EOS} with input from nuclear physics and astrophysics is commonly done with Bayesian inference~\cite{Raithel:2017ity, Radice:2017lry, Annala:2017llu, Raithel:2018ncd, Coughlin:2018fis, Capano:2019eae, Miller:2019nzo, Tews:2019cap, Li:2020bre, Legred:2021hdx, Huth:2021bsp, Biswas:2021pvm, Raaijmakers:2021uju, Pang:2021jta, Biswas:2021yge, Essick:2021ezp, Pang:2022rzc, Essick:2020flb, Brandes:2022nxa, Rose:2023uui, Brandes:2024wpq, Rutherford:2024srk, Biswas:2024hja, Huang:2024rfg, Raaijmakers2025, Kashyap:2025cpd, Wu:2025edc} (see Ref.~\cite{Koehn:2024set} for a recently compiled overview of \ac{EOS} constraints). 
However, current inference methods remain computationally expensive, particularly when directly sampling a set of \ac{EOS} parameters, as this requires solving the \ac{TOV} equations on the fly for millions of samples. 
This computational burden will hinder our ability to extract precise \ac{EOS} constraints from \ac{NS} data delivered by future instruments, such as upcoming X-ray telescopes (e.g., eXTP and STROBE-X~\cite{eXTP:2018anb, Watts:2018iom, STROBE-XScienceWorkingGroup:2019cyd}) and the next-generation of ground-based \ac{GW} detectors (i.e., the Einstein Telescope~\cite{Punturo:2010zza, ET:2019dnz, ETScienceCase2020, ETDesignUpdate2024} and Cosmic Explorer~\cite{Evans:2021gyd}).
For example, next-generation \ac{GW} observatories are projected to detect approximately $\mathcal{O}(10^5)$ \ac{BNS} mergers annually~\cite{ET:2019dnz, Piro:2021oaa}. 
These detectors will measure \ac{NS} radii potentially with an unprecedented accuracy of $10 - 100$ meters~\cite{Chatziioannou:2021tdi, Puecher:2023twf, Iacovelli:2023nbv, Huxford:2023qne, Walker:2024loo}.
This increase in both the quantity and quality of observations highlights the need for more efficient inference methods for the inverse problem of \acp{NS}.
Ideally, such methods should be capable of solving the \ac{TOV} equations on the fly to ensure accurate and robust \ac{EOS} constraints.

One possible solution, thoroughly explored in past works, is to resort to machine-learning surrogates.
For instance, previous works have accelerated the \ac{TOV} solver by replacing it with machine learning-based approximations using regression methods~\cite{Richter:2023zec, Imam:2023ngm}, neural networks~\cite{Ferreira:2019bny, Thete:2022drz, Tiwari:2024jui, Reed:2024urq, Magnall:2024ffd, DiClemente:2025pbl,Somasundaram:2024ykk}, support vector machines~\cite{Ferreira:2019bny}, Gaussian processes and reduced basis methods~\cite{Reed:2024urq}, and dynamic mode decomposition~\cite{Lalit:2024vmu} (see Ref.~\cite{Zhou:2023pti} for a review). 
Other works, rather than relying on traditional sampling techniques, use machine learning models to directly recover the \ac{EOS} from \ac{NS} data, e.g., with neural networks~\cite{Fujimoto:2017cdo, Fujimoto:2019hxv, Morawski:2020izm, Traversi:2020dho, Krastev:2021reh, Fujimoto:2021zas, Soma:2022qnv, Soma:2022vbb, Farrell:2022lfd, Han:2022sxt, Ferreira:2022nwh, Carvalho:2023ele, Soma:2023fyi, Krastev:2023fnh, Wu:2023npy, Chatterjee:2023ecc, Ventagli:2024xsh}, normalizing flows~\cite{Morawski:2022aud, Brandes:2024vhw, Hu:2024oen}, transformers~\cite{Goncalves:2022smd}, principal component analysis~\cite{Patra:2023jbz}, and symbolic regression~\cite{Imam:2024xqg, Patra:2025sak, Patra:2025xtd, Bejger:2025hbq}. 
However, training these surrogate models can take on the order of hours~\cite{McGinn:2024nkd} to weeks~\cite{Hu:2024oen} (depending on the model complexity and hardware used for training) and requires additional modeling efforts. 
Moreover, we are often interested in comparing inferences with different parameterizations or prior assumptions.
Since emulators must be retrained whenever these change, their computational speed-up comes at the cost of reduced flexibility in inference. 

In this work, we propose an alternative solution to efficiently solve the inverse problem of \acp{NS} without resorting to machine-learning emulators. 
In particular, we leverage differential programming and hardware acceleration with \acp{GPU} to infer the \ac{EOS} from \ac{NS} properties in two ways. 
First, we make Bayesian inference of the \ac{EOS} from \ac{NS} data in high-dimensional \ac{EOS} parameter spaces computationally efficient. 
Second, we introduce a gradient-based optimization scheme that can recover the parameters of the \ac{EOS} underpinning a given  \ac{NS} mass-radius curve.
While the former method can take observational uncertainties into account, the latter can be used to improve our understanding of how accurate knowledge of \acp{NS} propagates into constraints on the \ac{EOS}. 

This paper is organized as follows. 
In Sec.~\ref{sec:methods}, we provide an overview of the \ac{EOS} parametrization used in this work, and discuss Bayesian inference and our gradient-based optimization scheme.
In Sec.~\ref{sec:results}, we validate the results obtained with our Bayesian inference code and investigate its scaling as a function of the dimensionality of the \ac{EOS} parameter space.
Moreover, we demonstrate how our framework can deliver insights into the inverse mapping from \ac{NS} data to \ac{EOS} parameters.
Finally, we discuss our results in Sec.~\ref{sec:discussion} and conclude in Sec.~\ref{sec:conclusion}. 

\section{Methods}\label{sec:methods}

\subsection{Equation of state}\label{sec: methods EOS parametrization}

Several approaches exist to parametrize the \ac{EOS} without resorting to a specific physical description of supra-nuclear matter. 
These include piecewise polytropes~\cite{Read:2008iy,Steiner:2010fz, Hebeler:2013nza, Raithel:2016bux, OBoyle:2020qvf}, the spectral representation~\cite{Lindblom:2010bb, Lindblom:2012zi, Lindblom:2018rfr, Fasano:2019zwm, Lindblom:2022mkr}, a speed-of-sound parametrization~\cite{Tews:2018iwm, Greif:2018njt}, and non-parametric models based on Gaussian processes~\cite{Landry:2018prl, Essick:2019ldf, Landry:2020vaw} or neural networks~\cite{Han:2021kjx, Li:2025obt}. 
In this work, we adopt the parametrization of Ref.~\cite{Koehn:2024set}.

At the lowest densities, up to $0.5 n_{\rm{sat}}$, the crust \ac{EOS} is fixed to the model from Ref.~\cite{Douchin:2001sv}. 
At densities just above the crust-core transition, we assume that matter consists solely of nucleonic degrees of freedom in $\beta$-equilibrium and use a \ac{MM} parametrization of the \ac{EOS}~\cite{Margueron:2017eqc, Margueron:2017lup, Somasundaram:2020chb}.
This parametrizes the energy per nucleon 
\begin{align}
\begin{split}
    e(n, \delta) = e_{\rm{sat}}(n) + e_{\rm{sym}}(n) \delta^2  + \mathcal{O}(\delta^4) \, ,
\end{split}
\label{eq:energy_per_nucleon_MM}
\end{align}
where $n$ represents the baryon number density and ${\delta = (n_n - n_p) / n}$ is the asymmetry parameter, $n_n$ and $n_p$ denote the neutron number density and the proton number density, respectively.
The isoscalar (saturation) energy $e_{\rm{sat}}$ and the isovector (symmetry) energy $e_{\rm{sym}}$ are defined as a series expansion in $x = (n - n_{\rm{sat}}) / 3 n_{\rm{sat}}$. 
The Taylor expansion of the former is given by
\begin{align}
\begin{split}
    e_{\rm{sat}}(n) = E_{\rm{sat}} + \frac12 K_{\rm{sat}} x^2 + \frac{1}{3!} Q_{\rm{sat}} x^3 + \frac{1}{4!} Z_{\rm{sat}} x^4 + \dots \, ,
\end{split}
\label{eq: e_sat}
\end{align}
Throughout this work, we fix $E_{\rm{sat}} = -\SI{16}{\MeV}$ and $n_{\rm{sat}} = 0.16 \ \rm{fm}^{-3}$. The expansion of the symmetry energy is given by
\begin{align}
\begin{split}
    e_{\rm{sym}}(n) &= E_{\rm{sym}} + L_{\rm{sym}} x + \frac12 K_{\rm{sym}} x^2  \\
    &+ \frac{1}{3!} Q_{\rm{sym}} x^3 + \frac{1}{4!} Z_{\rm{sym}} x^4 + \dots \, .
\end{split}
\label{eq: e_sym}
\end{align}
The expansion coefficients in Eqs.~\eqref{eq: e_sat} and~\eqref{eq: e_sym} are referred to as the \acp{NEP}.  
The \ac{MM} parametrization allows us to incorporate constraints from nuclear-physics experiments or compare them with similar constraints obtained from astrophysical observations. 
For instance, the slope of the symmetry energy $L_{\rm{sym}}$ is correlated with both the neutron-skin thickness as well as the radii of \acp{NS}~\cite{Typel:2001lcw, Lattimer:2015nhk, Russotto:2023ari}. 

However, this parametrization, describing nucleonic degrees of freedom, is inadequate to model the higher-density regime probed by \acp{NS}, as new degrees of freedom might become relevant~\cite{Lattimer:2006qiu, Oertel:2016xsn, Kumar:2023lhv}, possibly with a phase transition~\cite{Baym:2017whm, Pang:2020ilf, Bogdanov:2022faf, Mondal:2023gbf, Essick:2023fso, Prakash:2023afe, Huang:2025vfl}. 
Therefore, we use a phenomenological and model-agnostic approach at higher densities.
Specifically, we use a \ac{CSE} parametrization above a breakdown density $n_{\rm{break}}$, which is varied freely in our parametrization within $1-2 \ \nsat$~\cite{Tews:2018iwm}, unless stated otherwise. 
At densities above $n_{\rm{break}}$, we parametrize the speed-of-sound $c_s$ using a non-uniform grid of points in the $(n, c_s^2)$-plane and construct the full speed-of-sound profile with linear interpolation.
We extend this \ac{CSE} to $25 \nsat$, the end-point of our \acp{EOS}, for which $c_s^2$ is also sampled. 
The speed-of-sound curve from the \ac{CSE} parametrization is then matched to that of the \ac{MM} parametrization with cubic splines. 
The \ac{EOS} can then be derived from the speed-of-sound curve, see Ref.~\cite{Somasundaram:2021clp}.

We refer to the parameters of our parametrization, encompassing the \acp{NEP} and the \ac{CSE} grid points, as $\thetaeos$. 

\subsection{Neutron star observables}

The macroscopic properties of non-spinning, isolated, spherically symmetric \acp{NS} are determined through the \ac{TOV} equations~\cite{Tolman:1939jz, Oppenheimer:1939ne}, the general-relativistic equations determining hydrostatic equilibrium\footnote{We refer readers to Ref.~\cite{glendenning2010special} for a didactic derivation of the \ac{TOV} equations.}:
\begin{align}
    \frac{\diff p}{\diff r} &= - \frac{\varepsilon(r)m(r)}{r^2}\frac{\left[ 1 + \frac{p(r)}{\varepsilon(r)} \right] \left[1 + \frac{4 \pi r^3 p(r)}{m(r)} \right]}{ \left[1 - \frac{2m(r)}{r}\right]} \, , \\ 
    \frac{\diff m}{\diff r} &= 4 \pi r^2 \varepsilon(r) \, ,
\end{align}
where $\varepsilon$ is the energy density, $r$ is the radial coordinate from the center of the \ac{NS} and $m(r)$ is the mass contained in a sphere of radius $r$. 
Here, we employ units where $G=c=1$. 
The system of equations is closed by the \ac{EOS}, which links pressure and energy density. 
Given a central pressure $p_c$ of the \ac{NS}, the \ac{TOV} equations can be integrated outwards, with boundary conditions ${m(r = 0) = 0}$ and ${p(r = 0) = p_c}$, until we reach the boundary of the star where ${p(r = R) = 0}$. 
This gives the star's radius $R$ and gravitational mass ${M \equiv m(R)}$. 
Repeating this procedure for various $p_c$ provides the mass-radius relation of \acp{NS} for that \ac{EOS}. 

When an \ac{NS} is placed in a static, external quadrupolar tidal field $\mathcal{E}_{ij}$, it develops a quadrupole moment $Q_{ij}$~\cite{Hinderer:2009ca, Damour:2009vw, Damour:2012yf}.
This induced deformation is described by the $\ell = 2$ eigenfunctions of the oscillations and, to linear order, takes the form $Q_{ij} = - \lambda \mathcal{E}_{ij}$, where $\lambda$ is related to the second tidal Love number $k_2$ through $k_2 = \frac32 \lambda R^{-5}$. 
We refer readers to Refs.~\cite{Hinderer:2009ca, Damour:2009vw, Damour:2012yf} for further details. 
These tidal deformations occur, for instance, in \ac{BNS} systems, where the gravitational field of each \ac{NS} induces the tidal effects deforming its companion. 
This leaves an imprint on the phase of the \acp{GW} emitted during inspiral. 
In particular, the \ac{GW} emission depends, at leading order, on each star's dimensionless tidal deformability parameter $\Lambda$, which is defined as
\begin{equation}
    \Lambda = \frac23 k_2 \frac{R^5}{M^5} \, .
\end{equation}

\subsection{Differentiable programming}\label{sec: methods autodiff}

Differentiable programming is a programming paradigm that enables the end-to-end differentiation of computer programs~\cite{Baydin:2015tfa, sapienza2024differentiableprogrammingdifferentialequations, blondel2024elementsdifferentiableprogramming}.
The derivative is computed with automatic differentiation, which exploits the fact that all numerical calculations are composed of elementary operations.
The derivatives of these operations are accumulated during runtime to efficiently evaluate the gradient of a program. 

As a result, the parameters used in a program can easily be optimized using, for instance, gradient descent. 
If the program expresses a function $\mathcal{L}(\boldtheta)$, then gradient descent iteratively adapts the parameters $\boldtheta$ along the gradient of the objective function:
\begin{equation}\label{eq: gradient descent}
    \boldtheta^{(i+1)} \leftarrow \boldtheta^{(i)} \pm \gamma \nabla \mathcal{L}(\boldtheta^{(i)}) \, ,
\end{equation}
where the hyperparameter $\gamma$ is often referred to as the learning rate and the plus sign (minus sign) is used when maximizing (minimizing) the objective function $\mathcal{L}$. 

In this work, we use \textsc{jax}~\cite{frostig2018compiling, kidger2021on}, which provides the aforementioned automatic differentiation capabilities in Python. 
Moreover, \textsc{jax} can compile Python code and execute it on hardware accelerators, such as a \ac{GPU}. 
To exploit these features, we have developed a Python package implementing \textsc{jax}-based \ac{EOS} code and a \ac{TOV} solver, called \textsc{jester}\footnote{\url{https://github.com/nuclear-multimessenger-astronomy/jester}}. 
With \textsc{jester}, we can efficiently evaluate the derivative of any numerical function that depends on the \ac{EOS} parameters, even if it involves solving the \ac{TOV} equations to obtain the macroscopic properties of the \ac{NS} as an intermediate step.

While the \ac{TOV} boundary condition (pressure reaching zero) appears non-differentiable, our implementation circumvents this challenge through two key techniques.
First, we parameterize the TOV equations using enthalpy $h$ as the integration variable (similar to, e.g., Ref.~\cite{Lindblom1992-zb}), which transforms the stopping condition from ${p(r) = 0}$ to the smooth endpoint ${h = 0}$.
Second, we pre-allocate arrays for a maximum number of integration steps and employ bounded while loops to populate the solution arrays, after which \textsc{jax} operations mask and truncate the results to extract only the physically meaningful portion.
A similar strategy is employed to truncate the \ac{NS} solutions at the \ac{TOV} mass.
Since these operations rely entirely on \textsc{jax} primitives, the entire code supports automatic differentiation.

In the following sections, we introduce how we leverage automatic differentiation in gradient-based samplers and a novel gradient-based inversion scheme for \acp{NS}. 

\subsection{Bayesian inference}\label{sec: methods Bayesian inference}

The inverse problem of \acp{NS} (i.e., inferring the \ac{EOS} from \ac{NS} data) is often tackled with Bayesian inference, where the goal is to infer the posterior distribution $P(\thetaeos | d)$ of the \ac{EOS} parameters given a dataset $d$, representing either \ac{NS} observations or constraints from nuclear physics. 
This is achieved by using Bayes' theorem
\begin{equation}\label{eq: Bayes theorem}
    P(\thetaeos | d) = \frac{P(d | \thetaeos) P(\thetaeos)}{P(d)} \, ,
\end{equation}
where $P(d | \thetaeos)$ is the likelihood function, $P(\thetaeos)$ the prior distribution and $P(d)$ the Bayesian evidence. 

Since the posterior distribution is in most practical applications analytically intractable, one has to sample it numerically with methods such as nested sampling~\cite{Skilling:2006gxv} or \ac{MCMC}~\cite{Neal:2011mrf}. 
In this work, we use \textsc{flowMC}~\cite{Gabrie:2021tlu, Wong:2022xvh}, which evolves Markov chains along the gradient of the likelihood toward high-likelihood regions~\cite{grenander1994representations}. 
Moreover, to improve the sampling efficiency, \acp{NF} are used as proposal distributions.
\Acp{NF} are a class of deep-learning models that offer tractable representations of complex probability distributions~\cite{Papamakarios:2019fms, kobyzev2020normalizing}.
During sampling, the \ac{NF} is trained on the fly from the available \ac{MCMC} samples, after which it is used as a global proposal distribution, improving the mixing of the chains. 

It has previously been shown that \textsc{flowMC} generates more effective samples than an \ac{MCMC} algorithm that only uses a gradient-based sampler~\cite{Gabrie:2021tlu}.
In this work, we use the Metropolis-adjusted Langevin algorithm (MALA) for the gradient-based sampler~\cite{grenander1994representations}.
We have not tested our setup with the NUTS sampler, a specific implementation of Hamiltonian Monte Carlo (HMC)~\cite{Betancourt:2017ebh, Hoffman:2011ukg}.
However, MALA is computationally cheaper to run (see, e.g., Ref.~\cite{Neural_Langevin_Dynamical_Sampling} for runtime comparisons on some numerical experiments) and gives satisfactory results in terms of effective sample size.

\subsection{Equation of state constraints}\label{sec: methods EOS constraints overview}

In this work, we consider constraints on the \ac{EOS} from \ac{chiEFT}, mass measurements of heavy pulsars, mass-radius constraints provided by the \ac{NICER} instrument, and the tidal deformabilities measured in GW170817. 
Below, we provide more details on each of these constraints.

\subsubsection{Chiral effective field theory}

At low energies, \ac{chiEFT} provides an expansion of the nuclear Hamiltonian in terms of the nucleon momenta over a breakdown scale~\cite{Epelbaum:2008ga, Machleidt:2011zz}.
After truncating the expansion at a certain order, the nuclear many-body problem involving the Hamiltonian can be solved numerically, from which the \ac{EOS} follows.
The truncation introduces theoretical uncertainty, which, however, can be systematically estimated~\cite{Epelbaum:2014efa, Drischler:2020hwi, Drischler:2020yad,Armstrong:2025tza}. 
As a result, \ac{chiEFT} provides an uncertainty band for the \ac{EOS}, bounded by two pressure curves $p_-(n)$ and $p_+(n)$.
We constrain the \ac{EOS} using this band, following an approach similar to Ref.~\cite{Koehn:2024set}.
In particular, the calculation done by Ref.~\cite{Koehn:2024set} yields a score function $f(p, n)$ that quantifies the consistency between a given \ac{EOS} and the \ac{chiEFT} prediction.
Given this score function, the likelihood is given by
\begin{equation}
    P(\thetaeos|d_{\chi \rm{EFT}}) \propto \exp\left( \int_{0.75 n_{\rm{sat}}}^{n_{\rm{break}}} \frac{\log f(p(\thetaeos ; n), n)}{n_{\rm{break}} - 0.75 n_{\rm{sat}}} \diff n \right) \, .
\end{equation}
We utilize the pressure band from Ref.~\cite{Tews:2018kmu}, which extended calculations for pure neutron matter to $\beta$-equilibrium, making it applicable to \ac{NS} \ac{EOS}.

The approach used in this work neglects the correlations between pressures.
However, we follow the same approach as Ref.~\cite{Koehn:2024set} in order to fully match their setup for our comparison in Sec.~\ref{sec: results validation and scaling}.

\subsubsection{Mass measurements of pulsars}

Mass measurements of the heaviest known pulsars, obtained by evaluating relativistic effects in radio timing, provide a lower bound for the maximal mass of an \ac{NS}, also referred to as the \ac{TOV} mass $\MTOV$. 
Here, we consider three massive pulsars: PSR J1614-2230~\cite{Demorest:2010bx, Shamohammadi:2022ttx} with a mass of $1.937 \pm 0.014 \ M_\odot$, PSR J0348+0432 with a mass of $2.01 \pm 0.04 \ M_\odot$~\cite{Antoniadis:2013pzd},~\footnote{Note that recent work~\cite{Saffer:2024tlb} argues the mass to be $\sim 1.8 M_\odot$. However, in this work, we stick to the mass estimate determined previously, to allow for a direct comparison to Ref.~\cite{Koehn:2024set}.} and PSR J0740+6620 with a mass of $2.08 \pm 0.07 \ M_\odot$ (but see also Refs.~\cite{NANOGrav:2019jur, NANOGrav:2023hde}).
All values are quoted at the $68\%$ credible level.
Radio timing measurements $d_{\rm{radio}}$ give a posterior on the pulsar mass $P(M | d_{\rm{radio}})$, which is well approximated by a normal distribution with the mean and standard deviation values quoted above.
For each pulsar, the likelihood on the \ac{EOS} parameters is then
\begin{equation}\label{eq: EOS posterior for radio timing}
    P(\thetaeos|d_{\rm{radio}}) \propto \frac{1}{\MTOV} \int_{0}^{\MTOV} P(M|d_{\rm{radio}}) \diff M \, ,
\end{equation}
where the \ac{EOS} dependence is implicit in $\MTOV$ and we assumed a uniform distribution on the masses. 

\subsubsection{NICER}

Pulse profile modeling of \ac{NICER} data has provided mass-radius posteriors for various pulsars~\cite{Riley:2019yda, Miller:2019cac, Riley:2021pdl, Miller:2021qha, Choudhury:2024xbk, Salmi:2024bss}.
In this work, we limit ourselves to those of PSR J0030+0451 and PSR J0740+6620. 
These analyses have later been refined with more accumulated data and advanced models for the hotspots and instrument response~\cite{Vinciguerra:2023qxq, Salmi:2024aum, Dittmann:2024mbo}. 
We use posterior samples on the masses and radii from the public data releases~\cite{NICER_posterior_samples_Zenodo_J0030_Riley, NICER_posterior_samples_Zenodo_J0030_Miller, NICER_posterior_samples_Zenodo_J0740_Salmi, NICER_posterior_samples_Zenodo_J0740_Dittmann}. 
Their distributions $P(M, R|d_{\rm{NICER}})$ are approximated by a Gaussian kernel density estimate.
In our inference, we sample the mass $M$ of each pulsar from a uniform prior distribution and compute the likelihood as
\begin{equation}\label{eq: EOS posterior for NICER}
    P(\boldsymbol{\theta} | d_{\rm{NICER}}) \propto P(M, R(\thetaeos; M)) | d_{\rm{NICER}}) \, ,
\end{equation}
where the parameters $\boldsymbol{\theta}$ consist of the \ac{EOS} parameters and the pulsar masses. 

\subsubsection{GW170817}

Finally, we consider constraints from tidal deformability measurements of GW170817.\footnote{The second observed \ac{BNS} candidate so far, GW190425~\cite{LIGOScientific:2020aai}, does not significantly constrain the \ac{EOS} due to its lower signal-to-noise ratio and larger component masses, and therefore will not be considered in this work.}
For this, one first has to determine the posterior distributions on the source parameters $\boldsymbol{\theta}_{\rm{GW}}$ of GW170817~\cite{Veitch:2009hd, Veitch:2014wba, Romero-Shaw:2020owr}.
Here, we take the posterior samples produced by the analysis of Ref.~\cite{Wouters:2024oxj}, which obtained the posterior using \textsc{Jim}~\cite{Edwards:2023sak, Wong:2023lgb, Wouters:2024oxj}, a \ac{GW} inference toolkit implemented in \textsc{jax} able to perform fast Bayesian inference on \ac{GW} signals. 
The samples are publicly available at Ref.~\cite{Jim_GW170817_posterior_samples}. 

From the posterior on the detector-frame chirp mass, the mass ratio, and the luminosity distance, we infer the source-frame component masses $M_i$ using the linear Hubble law, where $i = 1, 2$ labels the two components of the binary. 
For the inference on the \ac{EOS}, we then consider the marginalized posterior distribution on $M_i$ and the tidal deformabilities $\Lambda_i$. 
Contrary to the case of the \ac{NICER} mass-radius posteriors, a Gaussian kernel density estimate provides a poor approximation of this distribution.
Therefore, we train\footnote{Training this \ac{NF} takes less than $\lesssim15$ minutes on an NVIDIA A100 \ac{GPU} using \textsc{FlowJAX}~\cite{ward2023flowjax}.} an \ac{NF} to obtain a tractable and more accurate approximation of the marginal posterior density, which we denote by $P_{\rm{NF}}(M_i, \Lambda_i | d_{\rm{GW}})$.
During inference, we sample the binary component masses $M_i$ and use the likelihood function
\begin{equation}\label{eq: EOS posterior for GW170817}
    P(\boldsymbol{\theta} | d_{\rm{GW}}) \propto P_{\rm{NF}}(M_i, \Lambda_i(\thetaeos; M_i) | d_{\rm{GW}}) \, ,
\end{equation}
where $\boldsymbol{\theta}$ consists of the \ac{EOS} parameters and source-frame component masses of GW170817. 

\subsection{Gradient-based inversion}\label{sec: methods gradient based optimization}

To assess how detailed knowledge on \ac{NS} properties propagates to the \ac{EOS} and the parametrization we employ for it, we have to invert a complete mass-radius relation $R(M)$ or mass-tidal deformability relation $\Lambda(M)$.
Here, we show that such an inversion can be achieved by employing a gradient-based optimization approach for the \ac{EOS} parameters, as an alternative to Bayesian inference, as the latter can be quite computationally expensive. 

In practice, we consider $N=500$ masses $M_i$, evenly spaced in the range $[1.0 \ M_\odot, \MTOV]$, and the corresponding radii $\hat{R}_i$ and tidal deformabilities $\hat{\Lambda}_i$ of the ``target'' \ac{EOS}, denoted by a hat.
We have found that the results of the recovery do not significantly depend on the value of $N$ and the spacing of the masses. 
The lower bound on the mass is conservatively chosen below the lowest reliably observed \ac{NS} mass to date~\cite{Martinez:2015mya}~\footnote{Note that an \ac{NS} candidate with a mass of $\sim 0.77 M_\odot$ was proposed in Ref.~\cite{Doroshenko:2022nwp}, though.}. 
We then measure the distance between a sample $\thetaeos$, with corresponding values for the tidal deformabilities $\Lambda_i(\thetaeos)$ at $M_i$, and the target \ac{NS} values with the following loss function:
\begin{equation}\label{eq: loss function EOS doppelgangers}
    \mathcal{L}(\thetaeos) = \frac1N \sum_{i=1}^N \left| \frac{\Lambda_i(\thetaeos) - \hat{\Lambda}_i}{\hat{\Lambda}_i} \right| \, .
\end{equation}
We note that the loss function is based on mock data coming from the ``true'' \ac{EOS} rather than existing \ac{NS} observations, and therefore is introduced in this work as a proof of concept rather than a full replacement of Bayesian inference.

\begin{figure}[tpb]
    \centering
    \includegraphics[width=0.95\linewidth]{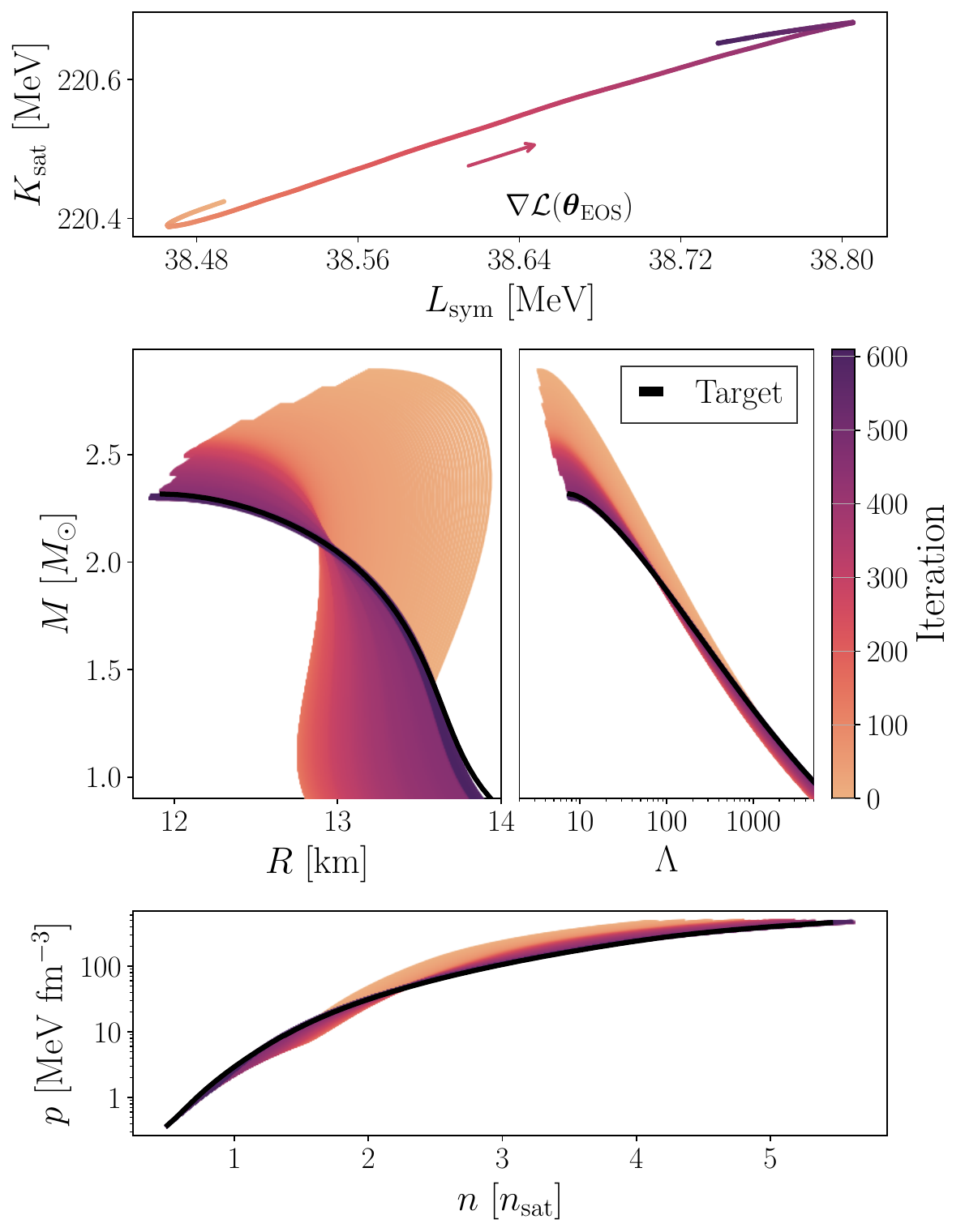}
    \caption{
    Illustration of the gradient-based inversion scheme recovering a given mass-radius relation of \acp{NS}. 
    The color gradient shows the different iterations while performing gradient descent on the objective function of Eq.~\eqref{eq: loss function EOS doppelgangers}.
    The gradient is computed with respect to the \ac{EOS} parameters throughout the \ac{TOV} equations using automatic differentiation.
    \textit{Top panel}: Trajectory of the \ac{EOS} parameters, showing only $L_{\rm{sym}}$ and $K_{\rm{sat}}$ for simplicity.
    \textit{Middle panel}: Mass-radius and mass-tidal deformability curves during the gradient descent, with the target shown in black.
    \textit{Lower panel}: Pressure as a function of density curves during optimization.
    }
    \label{fig: doppelganger trajectory for JESTER EOS}
\end{figure}

By starting from a randomized initial position in the \ac{EOS} parameter space and iteratively performing gradient descent on this loss function, as shown in Eq.~\eqref{eq: gradient descent}, we aim to recover a set of \ac{EOS} parameters that matches the given mass-tidal-deformability relation. 
The derivative of the loss function is computed with automatic differentiation, as explained in Sec.~\ref{sec: methods autodiff}.
In particular, we use the Adam optimizer, which stabilizes the descent with adaptive estimates of the first and second moments of the gradients~\cite{Kingma:2014vow}. 
Figure~\ref{fig: doppelganger trajectory for JESTER EOS} illustrates our optimization algorithm. 
The \ac{NS} curve of the \ac{EOS} at the final iteration deviates from the target \ac{NS} family with an error less than $100$ meters in radius and less than $10$ in tidal deformability for the mass range considered during optimization.
We note that we show the optimization run after truncating around $600$ iteration steps for visualization purposes.
We expect further improvement in the convergence, especially at lower masses, if the optimizer runs for more iterations.
Since the ``true'' \ac{EOS} considered in this example comes from the same \ac{EOS} parametrization as used in the recovery, a perfect recovery of the \ac{EOS} is theoretically possible.
However, in practice, the optimization can get stuck in local minima.
This might be more problematic in case the \ac{EOS} does not follow the same parametric form as used in the recovery.
Moreover, incorporating real data with uncertainties will change the loss function of Eq.~\eqref{eq: loss function EOS doppelgangers} and the robustness of the recovery procedure.
These additional complications will be addressed in future work.

While preparing our manuscript, Refs.~\cite{Lindblom:2025kjz, Li:2025obt} also introduced a recovery algorithm based on optimization. 
However, their methods differ from ours in several ways.
Ref.~\cite{Lindblom:2025kjz} uses a spectral expansion~\cite{Lindblom:2024yio} and relatively low-dimensional \ac{EOS} parameter spaces. 
Our scheme, instead, scales well to high-dimensional parameter spaces by employing gradient descent. 
Indeed, the illustration shown in Fig.~\ref{fig: doppelganger trajectory for JESTER EOS} performed the inversion on the complete \ac{EOS} parametrization described in Sec.~\ref{sec: methods EOS parametrization}, using eight grid points for the \ac{CSE}, and was able to recover the \ac{EOS}. 
Since this parametrization involves $26$ parameters in total, this example demonstrates the applicability of our method in high-dimensional \ac{EOS} spaces. 
Ref.~\cite{Li:2025obt} uses a gradient-based inversion scheme, but uses a neural network as \ac{EOS} parametrization and relies on linear response analysis to compute the gradients.
In contrast, we use automatic differentiation, which enables us to compute these gradients for any \ac{EOS} parametrization.

\section{Results}\label{sec:results}

\subsection{Validation and computational scaling}\label{sec: results validation and scaling}

We first demonstrate that our Bayesian inference methods reproduce previous results, and then investigate the scaling of the methodology as a function of the number of dimensions employed in the parametrization. 

In order to validate our method, we compare our results to those of Ref.~\cite{Koehn:2024set} which used an identical \ac{EOS} parametrization.\footnote{The quoted results for Ref.~\cite{Koehn:2024set} are obtained from \url{https://multi-messenger.physik.uni-potsdam.de/eos_constraints/}.} 
Moreover, we use the same prior distribution, given in Table~\ref{tab: prior distributions for the EOS parameters}. 
We take eight grid points for the \ac{CSE} parametrization. 

\begin{table}[tpb]
\centering
\caption{Table of \ac{EOS} parameters and their uniform prior distribution ranges, separated in the metamodel and speed-of-sound extension (CSE) parts, with the index $i$ labeling the grid points. Throughout this work, we fix $n_{\rm{sat}} = 0.16 \ \rm{fm}^{-3}$ and $E_{\rm{sat}} = -16$ MeV.}
\label{tab: prior distributions for the EOS parameters}
\renewcommand{\arraystretch}{1.35}
\begin{tabular*}{0.9\linewidth}{@{\extracolsep{\fill}} p{0.00001\linewidth} c c c}
\hline\hline
\multirow{8}{*}{\rotatebox{90}{Metamodel}} & Parameter & Prior \\
\hline
& $K_{\rm{sat}}$ [MeV] & $U(150, 300)$ \\
& $Q_{\rm{sat}}$ [MeV] & $U(-500, 1100)$ \\
& $Z_{\rm{sat}}$ [MeV] & $U(-2500, 1500)$ \\
& $E_{\rm{sym}}$ [MeV] & $U(28, 45)$ \\
& $L_{\rm{sym}}$ [MeV] & $U(10, 200)$ \\
& $K_{\rm{sym}}$ [MeV] & $U(-300, 100)$ \\
& $Q_{\rm{sym}}$ [MeV] & $U(-800, 800)$ \\
& $Z_{\rm{sym}}$ [MeV] & $U(-2500, 1500)$ \\ 
\hline
\multirow{3}{*}{\rotatebox{90}{CSE}} & $n_{\rm{break}}$ $[n_{\rm{sat}}]$ & $U(1, 2)$ \\
& $n^{(i)}$ $[n_{\rm{sat}}]$ & $U(n_{\rm{break}}, 25)$ \\ 
& $(c_s^{(i)}/c)^2$ & $U(0, 1)$ \\
\hline\hline
\end{tabular*}
\end{table}

In Table~\ref{tab: reproduction of Hauke}, we compare the $95\%$ credible intervals of the inferred radius of a $1.4 M_\odot$ \ac{NS}, $R_{1.4}$. 
The first rows of the table correspond to inferences employing the constraints discussed in Sec.~\ref{sec: methods EOS constraints overview} individually. 
The bottom row shows the result when considering all constraints jointly in a single Bayesian inference run. 
The table demonstrates that our results qualitatively agree with Ref.~\cite{Koehn:2024set}, but show some quantitative differences in the inferred radii.
Nevertheless, for all constraints considered here, we do observe quantitative agreement between both approaches for $\MTOV$, the pressure of the \ac{EOS} at $3\nsat$, and the maximal central density inside a \ac{NS}, $n_{\rm{TOV}}$, as shown by Table~\ref{tab: reproduction of Hauke appendix} in Appendix~\ref{app: further validation results}.

We attribute the greater impact on radii across the two studies to two main reasons.
First, the present work directly samples the \ac{EOS} parameters, whereas Ref.~\cite{Koehn:2024set} generated a fixed set of \ac{EOS} candidates which were weighted using the constraints.
Therefore, our approach enables a higher effective sample size and a more thorough exploration of the likelihood landscape. 
This would have the most impact on the inferred $R_{1.4}$ for the \ac{chiEFT}, \ac{NICER}, and \ac{GW} constraints, since their likelihoods directly constrain the \ac{EOS} at intermediate densities which affect $R_{1.4}$.
The radio timing constraint has less impact on $R_{1.4}$, as it mainly constrains the $\MTOV$, which is only weakly correlated with $R_{1.4}$ \textit{a priori}.
Second, while our implementation aims to match the \ac{EOS} parametrization of Ref.~\cite{Koehn:2024set} as closely as possible, there are small differences in our \textsc{jax} implementation (e.g., the interpolation in the crust-core transition) which translate into small differences in the implied prior distributions on $R_{1.4}$ when using the same priors from Tab.~\ref{tab: prior distributions for the EOS parameters}. 
Figure~\ref{fig: EOS and NS prior distributions} shows the prior distributions on the quantities shown in Tab.~\ref{tab: reproduction of Hauke} and Tab.~\ref{tab: reproduction of Hauke appendix}.
We observe a heavier tail towards larger radii in the prior on $R_{1.4}$ from this work, consistent with the larger inferred radii reported in Tab.~\ref{tab: reproduction of Hauke}.

\begin{table}[tpb]
\centering
\caption{Comparison of inferred $R_{1.4}$ ($95\%$ credible interval) in kilometers between Ref.~\cite{Koehn:2024set} and this work for different \ac{EOS} constraints. The bottom row refers to the posterior using all listed constraints jointly.}
\label{tab: reproduction of Hauke}
\renewcommand{\arraystretch}{1.35}
\begin{tabular*}{0.9\columnwidth}{@{\extracolsep{\fill}} l c c }
\hline\hline
& \multicolumn{2}{c}{$R_{1.4}$ [km]} \\  
Constraint & Ref.~\cite{Koehn:2024set} & This work \\
\hline
$\chi_{\rm{EFT}}$ & $12.11_{-3.39}^{+1.69}$ & $12.59_{-3.51}^{+2.24}$ \\
Radio timing      & $13.70_{-2.17}^{+1.41}$ & $13.71_{-1.88}^{+1.19}$ \\
PSR J0030+0451    & $13.17_{-2.24}^{+1.65}$ & $13.48_{-2.15}^{+1.42}$ \\
PSR J0740+6620    & $13.39_{-1.72}^{+1.57}$ & $13.79_{-1.73}^{+1.26}$ \\
GW170817          & $11.90_{-1.74}^{+1.78}$ & $12.24_{-2.18}^{+2.03}$ \\ \hline
All               & $12.26_{-0.91}^{+0.80}$ & $12.62_{-1.11}^{+0.97}$ \\
\hline\hline
\end{tabular*}
\end{table}

After validating that our framework agrees with previously published results, we now assess the runtime and its scaling as a function of the dimensionality of the \ac{EOS} parameter space.
Using a single NVIDIA H100 \ac{GPU}, the inferences listed in Table~\ref{tab: reproduction of Hauke} were performed under $1$ hour of wall time. 
Solving the \ac{TOV} equations to obtain the macroscopic properties of \acp{NS} ($100$ masses and their corresponding radii, and tidal deformabilities) for a single proposal takes around $\sim 0.3$ milliseconds, which is on par with previous works relying on machine learning surrogates (see, e.g., Refs~\cite{Reed:2024urq, Magnall:2024ffd, Tiwari:2024jui} for recent examples using neural networks), however, without the need of pretraining a machine learning model in our case. 

\begin{figure}[tpb]
    \centering
    \includegraphics[width=0.95\linewidth]{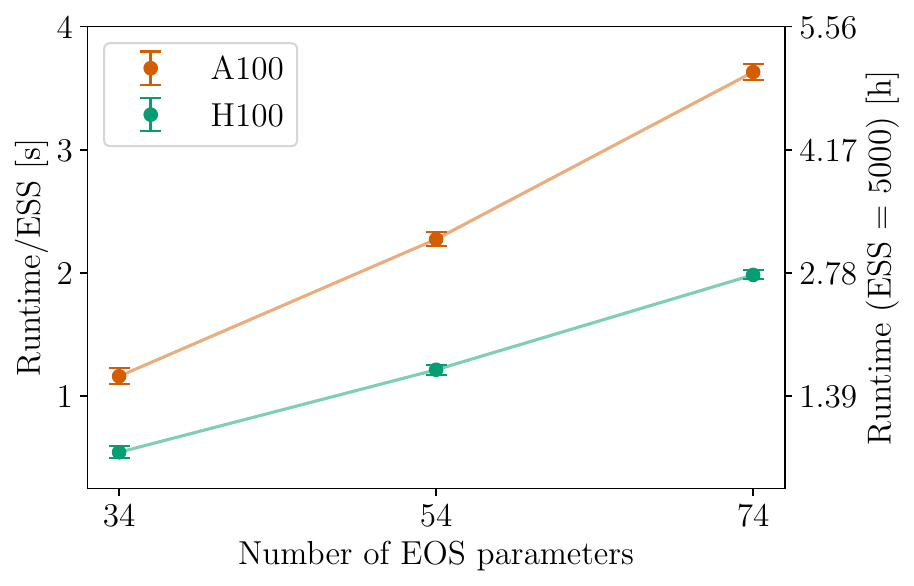}
    \caption{Scaling of the runtime divided by effective sample size (ESS) as a function of the number of \ac{EOS} parameters when varying the number of grid points for the \ac{CSE} parameterization, for two different \ac{GPU} architectures. The right axis converts this number to total wall time to obtain $5000$ effective samples.}
    \label{fig: GPU scaling plot}
\end{figure}

In the future, more precise observations of \acp{NS} will become available, which could require a more flexible \ac{EOS} parameterization to avoid biases.
This motivates us to extend the parameter space by adding more \ac{CSE} grid points and assess the impact on the scaling of our sampling efficiency.
Specifically, we perform Bayesian inference jointly on all \ac{EOS} constraints considered in this work by testing \ac{CSE} grids with $10$, $20$, and $30$ points, respectively.
Figure~\ref{fig: GPU scaling plot} shows the runtime per effective sample size (ESS), estimated with Geyer's initial monotone sequence criterion~\cite{geyer_SSE_ref1, geyer_SSE_ref2}. 
The axis on the right converts this number to an estimate of the total wall time required in case one requires $5000$ effective samples. 
We perform the scaling runs on two different \ac{GPU} architectures, namely an NVIDIA A100 and NVIDIA H100, and show the average and standard deviation over all the \ac{EOS} parameters. 
The parametrization with the highest number of grid points increases the runtime roughly by a factor $4$ to produce a similar number of effective samples.
Therefore, our inferences still only require a couple of hours to provide accurate posteriors on the \ac{EOS}, even when using a more flexible parametrization. 

\subsection{Constraining $\nbreak$ with neutron star data}\label{sec: measuring nbreak}

\begin{figure}[tpb]
    \centering
    \includegraphics[width=0.95\linewidth]{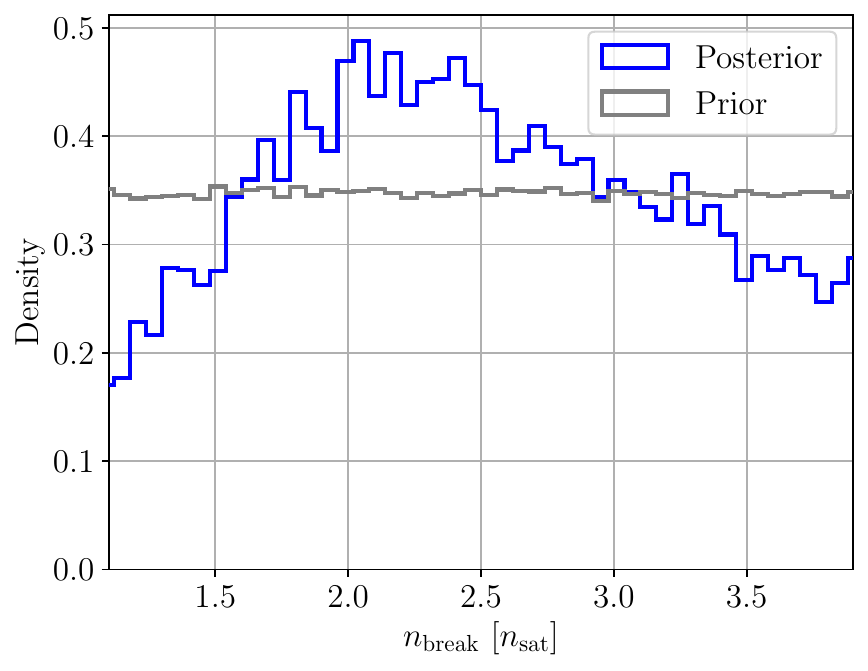}
    \caption{Posterior on $\nbreak$ of the metamodel parametrization inferred from a uniform prior between $1\nsat$ to $4\nsat$ (other priors are given in Tab.~\ref{tab: prior distributions for the EOS parameters}) and given the \ac{NS} observations from Tab.~\ref{tab: reproduction of Hauke}.}
    \label{fig: nbreak posterior}
\end{figure}

The \ac{EOS} parametrization employed in this work takes the breakdown density $\nbreak$ of the metamodel as a parameter that can be varied freely, after which the \ac{CSE} part of the parametrization is used for the high-density \ac{EOS}. 
This density is an indicator of the densities for which a simple nucleonic description of the EOS is not sufficient anymore.
We can use our framework to constrain $\nbreak$ using \ac{NS} data and directly probe this breakdown with astrophysical information. 
To this end, we perform Bayesian inference as explained above with two key modifications. 
First, we use a wider, agnostic prior on $\nbreak$, namely a uniform distribution between $[1 \nsat, 4 \nsat]$. 
The priors on the other \ac{MM} and \ac{CSE} parameters are identical to those shown in Tab.~\ref{tab: prior distributions for the EOS parameters}.
Second, we only consider \ac{EOS} constraints involving \ac{NS} data, i.e., we do not use the constraint from $\chiEFT$. 
Figure~\ref{fig: nbreak posterior} shows the marginal posterior distribution of $\nbreak$.
We infer a breakdown density of $\nbreak = 2.47_{-1.02}^{+1.53} \ \nsat$ at the $90\%$ credible level. 
Hence, \ac{NS} data prefers a smooth nucleonic \ac{EOS} up to 2-3 times the nuclear saturation density, a range within the reach of microscopic nuclear theory~\cite{Essick:2020flb}.
While $\nbreak$ is not tightly constrained by the data considered in this work, future detectors providing more \ac{NS} data can potentially improve its precision. 
Nevertheless, our result serves as proof of concept that the metamodel breakdown density can be determined on-the-fly with Bayesian inference. 

We point out that Ref.~\cite{Biswas:2020puz} similarly inferred this transition density. 
However, comparing the two results is complicated, as they use different parameterizations and prior choices.
In particular, we sample the $\nbreak$ parameter on the fly during inference, whereas  Ref.~\cite{Biswas:2020puz} infers this parameter by considering Bayes factors of different runs with fixed $\nbreak$. 
Moreover, we extend the \ac{MM} parametrization to higher orders and consider a speed-of-sound parametrization at high densities, rather than piecewise polytropes.

\subsection{Constraints on nuclear empirical parameters}\label{sec: results EOS degeneracy}

Currently available \ac{NS} observations do not yet tightly constrain the \ac{EOS}, as demonstrated by the previous section. 
In this section, we instead consider a hypothetical scenario where observations of \acp{NS} are pushed to extreme limits and uncertainties on the measured radii and tidal deformabilities can be ignored. 
We aim to obtain a qualitative estimate of how highly accurate knowledge of the properties of \acp{NS} translates into constraints on the \acp{NEP}, since such constraints can then be compared to those obtained from nuclear experiments. 
For simplicity, we restrict our parametrization to the metamodel part and no longer use the \ac{CSE} at high densities.
We enforce all these metamodel \acp{EOS} to be thermodynamically stable as well as to respect causality. 
Otherwise, the \acp{EOS} are truncated at the maximal central density allowed in an \ac{NS}.

We generate a mock ``true'' \ac{EOS}\footnote{Here, we did not consider agreement between this mock \ac{EOS} and the constraints from perturbative QCD at high densities, as this merely serves as a proof of concept of the methodology.} with a \ac{TOV} mass of roughly $2.0M_\odot$ and aim to recover its \ac{EOS} parameters from its $M$-$\Lambda$ curve using the gradient-based optimization scheme discussed in Sec.~\ref{sec: methods gradient based optimization}.
In particular, the inversion is performed with perfect knowledge of the $M$-$\Lambda$ curve, as discussed in detail in Sec.~\ref{sec: methods gradient based optimization}.
We leave a full study which considers Bayesian inference using the constraints from Tab.~\ref{tab: reproduction of Hauke}for future work.
We perform multiple such recoveries where we change the number of \acp{NEP} that are varied freely in the recovery, starting with only the first order parameter $L_{\rm{sym}}$. 
In subsequent recovery runs, we also allow the optimizer to explore the higher-order parameters, each time adding the parameters of the next order of the Taylor expansions of Eqs.~\eqref{eq: e_sat}-\eqref{eq: e_sym}.
For simplicity, we do not vary the $E_{\rm{sym}}$ parameter, since it has a small impact on the \ac{EOS} at high densities, and we wish to focus on the parameters entering at higher orders of the Taylor expansion. 
For each recovery, we start the optimizer from $100$ random \ac{EOS} samples, sampled from the priors shown in Table~\ref{tab: prior distributions for the EOS parameters}, and perform at most $1000$ iterations with a fixed learning rate of $1$.\footnote{The runs are performed on an AMD EPYC 7543 VM CPU and take at most a few minutes per recovery.}
For the final result, we keep only those \ac{EOS} samples that within the $[1\ M_\odot, 2\ M_\odot]$ mass range deviate less than $100$ meters in radius from the ``true'' $M$-$R$ curve and less than $10$ in tidal deformability from the ``true'' $M$-$\Lambda$ curve.

\begin{figure*}[tpb]
    \centering
    \includegraphics[width=\textwidth]{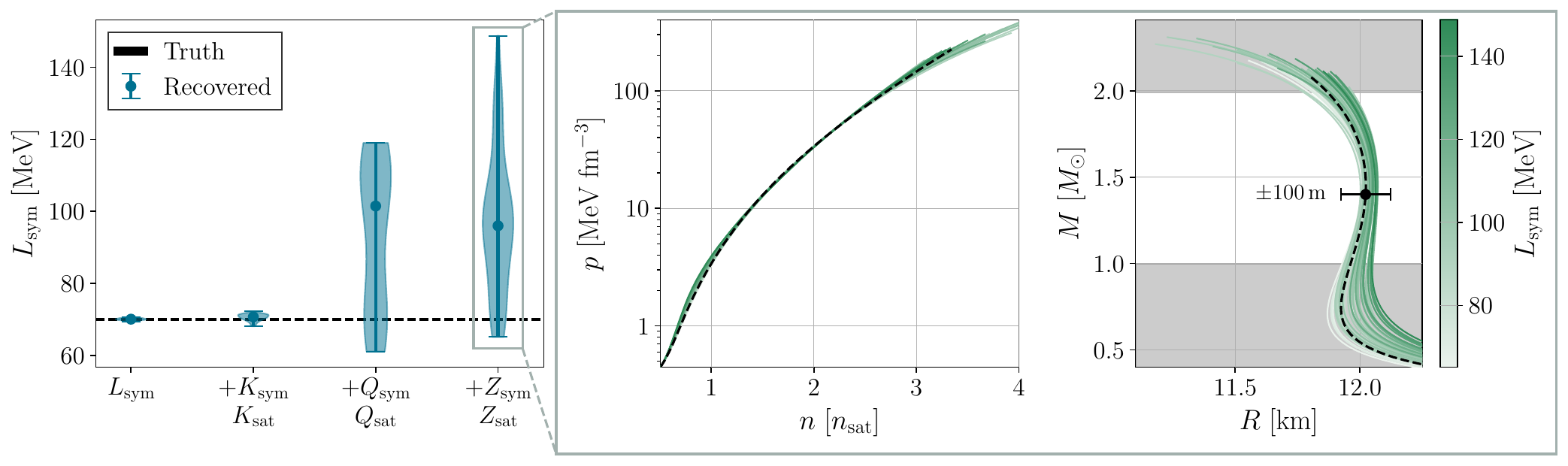}
    \caption{Recovery of an \ac{EOS} in the metamodel parametrization from the $M$-$\Lambda$ curve using the gradient-based optimization algorithm as a function of the number of \acp{NEP} being varied. 
    The left panel shows the recovered range of $L_{\rm{sym}}$ when increasingly more \acp{NEP} at higher orders in the Taylor expansion are varied freely. 
    For the recovery in which all \acp{NEP} were varied freely, we show the recovered \ac{EOS} samples in the middle and right panels.
    The middle panel shows the pressure as a function of density, while the right panel shows the mass-radius profile of \acp{NS}, color-coded by their $L_{\rm{sym}}$ value. 
    The grey bands show mass ranges not considered by the loss function in Eq.~\eqref{eq: loss function EOS doppelgangers}. 
    The \acp{EOS} shown in this figure deviate from the target by less than $100$ meters in radius and less than $10$ in dimensionless tidal deformability across the mass range considered in the loss function of Eq.~\eqref{eq: loss function EOS doppelgangers}.
    }
    \label{fig: results gradient based optimization}
\end{figure*}

Figure~\ref{fig: results gradient based optimization} shows the slope of the symmetry energy $L_{\rm{sym}}$ for these recovered \acp{EOS}. 
The range of the recovered $L_{\rm{sym}}$ values significantly depends on the number of \acp{NEP} that are allowed to vary. 
When $L_{\rm{sym}}$ is the only degree of freedom, we recover the value of the ``true'' \ac{EOS},  with a small width of around $0.5 \ \rm{MeV}$, as expected. 
However, when more \acp{NEP} are varied, this range increases. 
Nevertheless, the middle and right panels of Fig.~\ref{fig: results gradient based optimization} demonstrate (for the case where all \acp{NEP} are free parameters) that the corresponding \acp{EOS} and \ac{NS} configurations agree well with the true \ac{EOS}.
Therefore, from our gradient-based optimization approach, we conclude that, while accurate observations of \acp{NS} allow us to tightly constrain the \ac{EOS} itself, the corresponding constraints on the \acp{NEP} suffer from a degeneracy in our metamodel parametrization, which was already hinted at by Refs.~\cite{Krastev:2018nwr, Xie:2019sqb, Xie:2020tdo, Mondal:2021vzt, Iacovelli:2023nbv}. 
Therefore, fitting relations between $L_{\rm{sym}}$ and $R_{1.4}$ previously proposed in the literature, e.g., Ref.~\cite{Raithel:2020vvg}, will need to be assessed carefully in light of the assumptions regarding the parametrization used to obtain them. 
Finally, we note that this degeneracy is expected to be more pronounced if we include the \ac{CSE} parametrization at higher densities, as this would further diminish the impact of the individual \acp{NEP} at the higher-density regime of the \ac{EOS} to which \acp{NS} are most sensitive.

\section{Discussion}\label{sec:discussion}

\subsection{Equation of state degeneracies}

Our results from Sec.~\ref{sec: results EOS degeneracy} show that the inverse problem of \acp{NS} is sensitive to degeneracy in our metamodel parametrization. 
While this was shown using our gradient-based optimization scheme, future work will assess if these degeneracies remain when using traditional stochastic sampling algorithms.
A similar degeneracy was already observed in previous works that used stochastic samplers, e.g., Refs.~\cite{Krastev:2018nwr, Xie:2019sqb, Xie:2020tdo, Mondal:2021vzt, Iacovelli:2023nbv}. 
However, our work extends these studies by simultaneously considering (i) higher-order \acp{NEP}, (ii) a wide range of \ac{NS} masses, (iii) high accuracy in the recovered \acp{EOS}, highlighting the problem's relevance for future detectors, and (iv) the dependence as a function of the parametrization. 
This degeneracy is expected since the \ac{MM} expansion from Eq.~\eqref{eq: e_sat} - \eqref{eq: e_sym} is performed at $n = \nsat$, whereas \acp{NS} observations probe the \ac{EOS} at densities of a few $\nsat$, hinting that other parameterizations which avoid a Taylor expansion at a specific density could potentially alleviate this issue.
While our analysis in Sec.~\ref{sec: results EOS degeneracy} served as a theoretical exploration of this degeneracy, we expect that a full Bayesian analysis employing future constraints from nuclear theory and experiments on the \acp{NEP}, especially for the higher-order parameters of the \ac{MM}, alleviates this issue.

We expect the level of degeneracy to be different for different EOS models, and we will explore such model dependencies in future work. 
Furthermore, we will also investigate the potential introduction of similar systematic effects originating from certain aspects neglected in this study, such as the crust treatment (see, e.g., Refs.~\cite{Fortin:2016hny, Gamba:2019kwu, Ferreira:2020wsf, Rather:2020gja, Suleiman:2021hre, Canullan-Pascual:2025smm}), and consider unified \ac{EOS} models~\cite{Davis:2024nda} as alternative. 

\subsection{Next-generation detectors}

Future detectors will generate an unprecedented volume of \ac{NS} observations. 
Each event provides valuable constraints on the \ac{EOS}, yet, when analyzed with existing methods, at substantial economic and environmental costs~\cite{Hu:2024mvn}. 
Our work addresses this pressing data analysis challenge by employing differentiable programming and \ac{GPU} accelerators.
Using \textsc{jax}, analyzing a \ac{GW} signal from a \ac{BNS} merger to obtain masses and tidal deformabilities can be completed within minutes~\cite{Wouters:2024oxj}, after which we obtain posteriors on the \ac{EOS} in less than $1$ hour on an NVIDIA H100 \ac{GPU}. 
Crucially, we do not require pre-trained machine-learning emulators of the \ac{TOV} equations, thereby ensuring accuracy in the results and further reducing computational costs.
Therefore, our methods are a key advancement towards handling the data analysis problems offered by future detectors. 

Furthermore, while our constraint on the $\nbreak$ parameter from Sec.~\ref{sec: measuring nbreak} serves as a proof of principle, we expect that additional observations of \acp{NS} will tighten this constraint and allow us to test the consistency between nuclear physics and astrophysics.
Future research using simulated data will have to investigate to what accuracy the breakdown density of the metamodel can in fact be constrained from \ac{NS} data.
Moreover, the proof of concept presented here can be extended to directly test predictions from \ac{chiEFT}, providing an analysis complementary to Ref.~\cite{Essick:2020flb} with an alternative \ac{EOS} parametrization to assess possible systematic effects in such an analysis. 

\subsection{Extension to other analyses}

In this work, we only focused on inferring the \ac{EOS} from \ac{NS} observations.
However, other inferences require jointly modeling the \ac{EOS} together with additional effects or parameters.
Naturally, these analyses can therefore benefit from the acceleration offered by the methods presented in this work.

For instance, we can infer the \ac{EOS} simultaneously with cosmological parameters or population models (see, e.g., Refs.~\cite{Ghosh:2024cwc, Magnall:2024ffd, Golomb:2024lds}).
Moreover, our method can be extended to jointly infer the parameters of modified gravity, either for a specific theory (see Ref.~\cite{Olmo:2019flu} for an overview) or for a parametrized post-\ac{TOV} formalism~\cite{Glampedakis:2015sua, Glampedakis:2016pes, daSilva:2024qex}.
Our gradient-based scheme can be used to explore this extended parameter space and uncover degeneracies between the \ac{EOS} and deviations from general relativity in observations of \acp{NS}~\cite{Brown:2024joy}. 
Similar degeneracies can arise from other effects, such as phase transitions~\cite{Raithel:2022aee, Raithel:2022efm, Counsell:2025hcv} or dark matter~\cite{Koehn:2024gal}, which can also be constrained with either Bayesian inference or the gradient-based method. 

Going beyond extreme matter studies, other analyses in related fields that have to solve ordinary differential equations on the fly during inference can benefit from the methods discussed in this work, such as the evaluation of effective one-body gravitational waveforms~\cite{Damour:2009zoi},  solving the lens equation in strong lensing analyses (see, e.g., Refs.~\cite{Grespan:2023cpa, Poon:2024zxn}), and nuclear reaction networks~\cite{Reichert:2023xqy}. 

\acresetall

\section{Conclusion}\label{sec:conclusion}

The inverse problem of \acp{NS}, i.e., inferring the nuclear \ac{EOS} from \ac{NS} data, is a computationally challenging task, making it difficult to understand the interplay between nuclear physics and astrophysics.
Moreover, the prospect of a large volume of \ac{NS} observations delivered by future detectors calls for an efficient and scalable way of addressing this inverse problem. 
In this work, we have presented a solution based on differentiable programming, leveraging the automatic differentiation capabilities and support for hardware accelerators, such as \acp{GPU}, from \textsc{jax}. 
By using gradient-based samplers, we can perform \ac{MCMC} sampling on high-dimensional \ac{EOS} parameter spaces in under an hour of wall time on a single NVIDIA H100 \ac{GPU} without resorting to machine learning emulators for the \ac{TOV} equations.
Moreover, we provide a proof of concept on how the breakdown of the metamodel description can be inferred from \ac{NS} data on the fly during sampling. 

Additionally, our framework introduces a novel optimization scheme that recovers the \ac{EOS} of a given mass-radius or mass-tidal deformability relation by performing gradient descent on a loss function formulated in terms of the \ac{EOS} parameters, differentiating through the \ac{TOV} equations. 
Using this tool, we have discussed a degeneracy present in our metamodel parametrization, which might affect inferences of nuclear empirical parameters from detailed knowledge of \acp{NS} using certain \ac{EOS} parametrizations.
This optimization scheme offers a new way to investigate the relationship between the \ac{EOS} and the properties of \acp{NS}, and future work will investigate its robustness when the parametrization of the \ac{EOS} is misspecified or noisy observations are used.

In short, our methods leveraging differentiable programming address the data analysis challenges of future detectors and offer a pragmatic way to investigate the relation between \ac{EOS} parameterizations and properties of \acp{NS}. 

\appendix

\section{Method validations}\label{app: further validation results}

In Table~\ref{tab: reproduction of Hauke appendix}, we extend the results of Table~\ref{tab: reproduction of Hauke} and show $\MTOV$, the pressure at $3\nsat$, $p(3\nsat)$, and the maximal central density $n_{\rm{TOV}}$ at the core of an \ac{NS}, between Ref.~\cite{Koehn:2024set} and this work for the \ac{EOS} constraints discussed in Sec.~\ref{sec: methods EOS constraints overview}. 

\begin{table*}[tpb]
\centering
\caption{Comparison of inferred $\MTOV$, the pressure at $3\nsat$, $p(3\nsat)$, and the maximal central density $n_{\rm{TOV}}$ at the core of an \ac{NS}, between Ref.~\cite{Koehn:2024set} and this work for various \ac{EOS} constraints.
Uncertainties are quoted at the $95\%$ credible level.}
\label{tab: reproduction of Hauke appendix}
\renewcommand{\arraystretch}{1.35}
\begin{tabular*}{0.95\textwidth}{@{\extracolsep{\fill}} l c c c c c c}
\hline\hline
Constraint & \multicolumn{2}{c}{$\MTOV$ [$M_\odot$]} & \multicolumn{2}{c}{$p(3n_{\rm{sat}})$ [MeV fm${}^{-3}$]} & \multicolumn{2}{c}{$n_{\rm{TOV}}$ [$n_{\rm{sat}}$]} \\
& Ref.~\cite{Koehn:2024set} & This work & Ref.~\cite{Koehn:2024set} & This work & Ref.~\cite{Koehn:2024set} & This work \\
\hline
$\chi_{\rm{EFT}}$ & $2.05_{-1.16}^{+1.08}$ & $2.03_{-0.97}^{+1.03}$ & $\phantom{0}69_{-53}^{+186}$ & $\phantom{0}72_{-65}^{+165}$ & $6.51_{-3.11}^{+10.7}$ & $6.53_{-4.09}^{+9.26}$ \\
Radio timing      & $2.35_{-0.29}^{+0.73}$ & $2.20_{-0.26}^{+0.39}$ & $111_{-49}^{+140}$ & $\phantom{0}97_{-49\phantom{0}}^{+65\phantom{0}}$ & $5.51_{-1.66}^{+1.89}$ & $5.68_{-1.74}^{+1.63}$ \\
PSR J0030+0451    & $2.16_{-0.71}^{+0.83}$ & $2.19_{-0.76}^{+0.78}$ & $\phantom{0}89_{-46}^{+143}$ & $\phantom{0}94_{-62}^{+135}$ & $5.62_{-1.91}^{+4.44}$ & $5.60_{-2.59}^{+3.65}$ \\
PSR J0740+6620    & $2.34_{-0.32}^{+0.65}$ & $2.38_{-0.42}^{+0.70}$ & $107_{-40}^{+125}$ & $118_{-59}^{+150}$ & $5.34_{-1.61}^{+1.63}$ & $5.16_{-2.04}^{+1.72}$ \\
GW170817          & $1.90_{-0.41}^{+0.71}$ & $1.93_{-0.48}^{+0.73}$ & $\phantom{0}59_{-25\phantom{0}}^{+91\phantom{0}}$ & $\phantom{0}62_{-42\phantom{0}}^{+109}$ & $7.37_{-2.72}^{+3.64}$ & $7.24_{-3.21}^{+3.77}$ \\ \hline
All               & $2.25_{-0.22}^{+0.42}$ & $2.23_{-0.23}^{+0.41}$ & $\phantom{0}90_{-31\phantom{0}}^{+71\phantom{0}}$ & $\phantom{0}92_{-42\phantom{0}}^{+70\phantom{0}}$ & $5.92_{-1.38}^{+1.35}$ & $5.92_{-1.49}^{+1.38}$ \\
\hline\hline
\end{tabular*}
\end{table*}

\begin{figure*}
    \centering
    \includegraphics[width=0.95\linewidth]{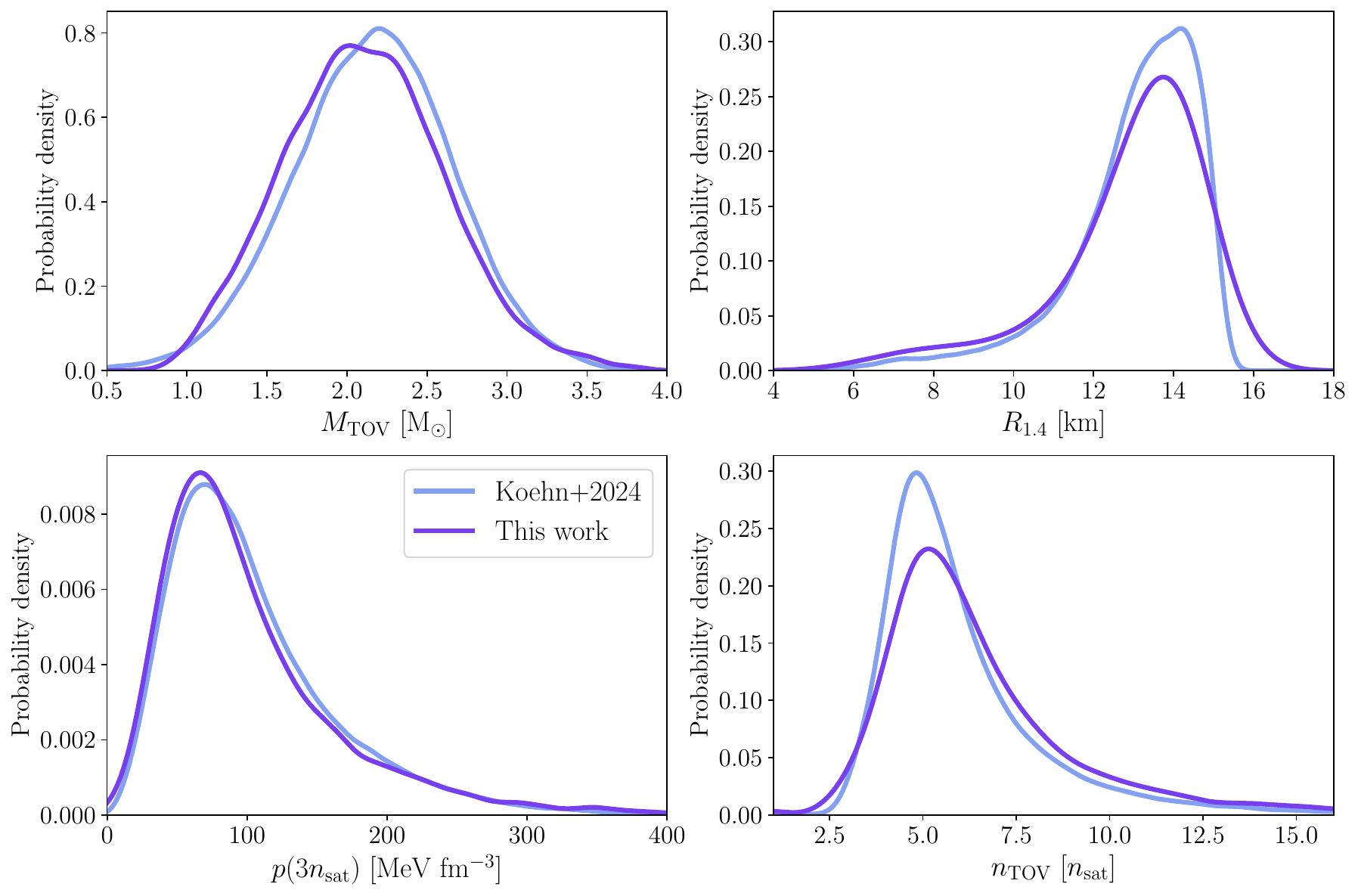}
    \caption{Comparison of prior distributions between Ref.~\cite{Koehn:2024set} (Koehn+2024) and this work on the \ac{EOS} quantities shown in Tab.~\ref{tab: reproduction of Hauke} and Tab.~\ref{tab: reproduction of Hauke appendix}.
    Since both use the \ac{EOS} parametrization described in Sec.~\ref{sec: methods EOS parametrization}, with priors given by Tab.~\ref{tab: prior distributions for the EOS parameters}, remaining differences are due to our \textsc{jax} implementation.
    }
    \label{fig: EOS and NS prior distributions}
\end{figure*}

\vspace{-2mm}

\section*{Data availability}

Our EOS code and TOV solver are available open source at \href{https://github.com/nuclear-multimessenger-astronomy/jester}{{\faGithub \texttt{nuclear-multimessenger-astronomy/jester}}}.
The code used in this study is available at Ref.~\cite{Github_repo_paper}.

\vspace{2mm}

\section*{Acknowledgments}

We thank Frank Snijder, Fabian Gittins, Harsh Narola, Justin Janquart, Anna Watts, Mariska Hoogkamer, Michael Coughlin, Micaela Oertel, and the LIGO-Virgo-Kagra extreme matter community for fruitful discussions and feedback that led to the improvement of this work.
T.W. and C.V.D.B. are supported by the research program of the Netherlands Organization for Scientific Research (NWO) under grant number OCENW.XL21.XL21.038.
P.T.H.P. is supported by the research program of the Netherlands Organization for Scientific Research (NWO) under grant number VI.Veni.232.021.
T.W., P.T.H.P., H.K., H.R., T.D.\ acknowledge support from the Daimler and Benz Foundation for the project ``NUMANJI''. T.D., H.K., H.R.\ acknowledges support from the European Union (ERC, SMArt, 101076369). 
Views and opinions expressed are those of the authors only and do not necessarily reflect those of the European Union or the European Research Council. Neither the European Union nor the granting authority can be held responsible for them.
R.S.\ acknowledges support from the Nuclear Physics from Multi-Messenger Mergers (NP3M) Focused Research Hub which is funded by the National Science Foundation under Grant Number 21-16686.
R.S.\ and I.T.\ were supported by the U.S. Department of Energy through the Los Alamos National Laboratory. 
Los Alamos National Laboratory is operated by Triad National Security, LLC, for the National Nuclear Security Administration of U.S. Department of Energy (Contract No.~89233218CNA000001).
R.S. and I.T.\ were also supported by the U.S. Department of Energy, Office of Science, Office of Advanced Scientific Computing Research, Scientific Discovery through Advanced Computing (SciDAC) NUCLEI program, and by the Laboratory Directed Research and Development Program of Los Alamos National Laboratory under project number 20230315ER.
We thank SURF (www.surf.nl) for the support in using the National Supercomputer Snellius under project numbers EINF-6587 and EINF-8596.
The authors acknowledge the computational resources provided by the LIGO Laboratory's CIT cluster, which is supported by National Science Foundation Grants PHY-0757058 and PHY-0823459.

\begin{acronym}
    \acro{AD}[AD]{automatic differentiation}
    \acro{JIT}[JIT]{just-in-time}
    \acro{PE}[PE]{parameter estimation}
    \acro{MCMC}[MCMC]{Markov chain Monte Carlo}
    \acro{GW}[GW]{gravitational wave}
    \acrodefplural{GWs}{gravitational waves}
    \acro{EM}[EM]{electromagnetic}
    \acro{CBC}[CBC]{compact binary coalescences}
    \acro{NS}[NS]{neutron star}
    \acrodefplural{NSs}{neutron stars}
    \acro{KDE}[KDE]{kernel density estimate}
    \acro{NF}[NF]{normalizing flow}
    \acro{BBH}[BBH]{binary black hole}
    \acro{BNS}[BNS]{binary neutron star}
    \acro{NSBH}[NSBH]{neutron star-black hole}
    \acro{EOS}[EOS]{equation of state}
    \acro{EFT}[EFT]{effective field theory}
    \acro{chiEFT}[$\chi$EFT]{chiral effective field theory}
    \acro{NEP}[NEP]{nuclear empirical parameter}
    \acro{HIC}[HIC]{heavy-ion collision}
    \acrodefplural{NEPs}{nuclear empirical parameters}
    \acro{MM}[MM]{metamodel}
    \acro{CSE}[CSE]{speed-of-sound extension scheme}
    \acro{TOV}[TOV]{Tolman-Oppenheimer-Volkoff}
    \acro{JS}[JS]{Jensen-Shannon}
    \acro{CPU}[CPU]{central processing unit}
    \acro{GPU}[GPU]{graphical processing unit}
    \acro{TPU}[TPU]{tensor processing unit}
    \acro{ML}[ML]{machine learning}
    \acro{SNR}[SNR]{signal-to-noise ratio}
    \acro{PSD}[PSD]{power spectral density}
    \acro{NICER}[NICER]{Neutron star Interior Composition ExploreR}
\end{acronym}

\bibliography{references}{}
\bibliographystyle{apsrev4-1}

\end{document}